# Decomposition of solid solutions in the Nd-Ba-Cu-O system and its influence on changing transport characteristics

N.I. Matskevich, O.I. Anyfrieva, T.D. Karpova *(Nikolaev Institute of Inorganic Chemistry, Siberian Branch of the Russian Academy of Science, Novosibirsk, Russia)*, Th. Wolf *(Institute of Solid State Physics, Karlsruhe, Germany)*, Eu.A. Trofimenko, Yu.D. Tretyakov *(Chemistry Faculty of the Moscow State University, Moscow, Russia)*

E-mail: nata@casper.che.nsk.su

**Abstract.** Thermochemical characteristics of $Nd_{1+x}Ba_{2-x}Cu_3O_y$ solid solutions were studied by solution calorimetry. Dependences of formation enthalpies from Nd content were constructed. It was established that decomposition of solid solution on the basis of Nd123 phase took place both in inert and in oxygen atmosphere. As a result of decomposition the different mixtures such as solid solutions with increasing and decreasing Nd content and $BaCuO_2$ were formed. Obtained thermochemical data were compared with results of measurements of transport characteristics. It was established that treatment of samples at 550-600° C led to increasing critical current and decreasing $T_c$.

**Introduction**

Development of technical application of high technical superconductors requires increasing the critical current density and irreversibility fields at 77 K. The best system to have large transition temperature and $J_c$ in comparison with YBCO is the Nd-Ba-Cu-O system. The

NdBa$_2$Cu$_3$O$_y$ (Nd123) superconductor displays enhanced flux pinning in high magnetic fields due to the anomalous peak-effect phenomenon [1-3]. This phenomenon is caused by nanoscale compositional fluctuation since Nd123 admits a wide solid solution range Nd$_{1+x}$Ba$_{2-x}$Cu$_3$O$_y$ (Nd123ss, x = 0.0-1.0 [1-4]), and the fluctuations appear as inhomogeneities due to solid solution demixing. At the same time the existence of solid solutions makes more difficult to control the physical properties of Nd123 based materials, such as decreasing $T_c$ values. Thus, there is a problem to create high quality NdBCO samples with high critical current. As it was described earlier [1, 5-8] there was no peak effect in perfect crystals and its appearance was connected with defects, such as twin boundaries, fine inclusions of non-superconducting phases, different types of microscopic and macroscopic cracks, etc. In order to achieve a better understanding of peak effect origins for the REBCO samples, it is necessary to investigate physico-chemical properties in the RE-Ba-Cu-O system.

As it was described in paper [9], it is possible that Nd-rich solid solutions should decompose into other phases. Authors [10] also reported that the Nd/Ba substitution can be achieved by a spinodal decomposition that can take place between 700-500° C. Decomposition plays important role in change of critical current. But there is no direct thermochemical investigation of solid solution decomposition in literature. There is no information about phases, which can be obtained during the decomposition process.

The aim of this paper is to investigate thermochemistry of solid solutions decomposition in the Nd-Ba-Cu-O system and to understand influence of this decomposition on changing critical current.

**Experimental section**

Nd$_{1+x}$Ba$_{2-x}$Cu$_3$O$_y$ ceramic samples were synthesized by solid state reaction using high-quality carbon-free precursors (Nd$_2$O$_3$, CuO and Ba(NO$_3$)$_2$). The samples with x from 0 up to 0.8 were synthesized. Other details see in paper [11, 12].

Nd$_{1+x}$Ba$_{2-x}$Cu$_3$O$_y$ single crystals were grown from the CuO-BaO flux in ZrO$_2$/Y crucible by the slow cooling method. They were subsequently oxidized several times between 600-300° C in an atmosphere of 1-400 bar O$_2$. These crystals are pure and stoichiometric with the exception of a small doping by Y. Other details see in paper [10].

All compounds were characterized by X-ray power diffraction and chemical analysis. Within the resolution of the analyses, the involved compounds were found to be single phase.

Solution calorimetry was used as investigated method. Technique and procedure of experiments were described in papers [11, 13].

**Results and discussion**

Performed calorimetric experiments allowed us to calculate the enthalpies of following reaction:

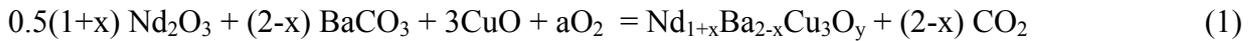

$$0.5(1+x)\ Nd_2O_3 + (2-x)\ BaCO_3 + 3CuO + aO_2 = Nd_{1+x}Ba_{2-x}Cu_3O_y + (2-x)\ CO_2 \qquad (1)$$

The obtained values allowed us to calculate formation enthalpies from oxides for the Nd$_{1+x}$Ba$_{2-x}$Cu$_3$O$_y$. Below the graphics of formation enthalpies dependences of obtained ceramic samples of Nd$_{1+x}$Ba$_{2-x}$Cu$_3$O$_y$ solid solutions are presented.

Fig. 1 shows the dependences of formation enthalpies of Nd$_{1+x}$Ba$_{2-x}$Cu$_3$O$_y$ solid solutions from Nd content. It can be seen that the decomposition of solid solutions in inert atmosphere in the interval from x = 0 up to x = 0.6 take place. The following reactions can proceed:

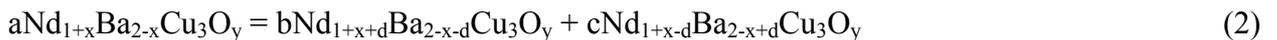

$$aNd_{1+x}Ba_{2-x}Cu_3O_y = bNd_{1+x+d}Ba_{2-x-d}Cu_3O_y + cNd_{1+x-d}Ba_{2-x+d}Cu_3O_y \qquad (2)$$

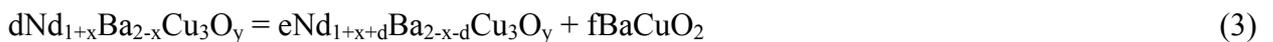

$$dNd_{1+x}Ba_{2-x}Cu_3O_y = eNd_{1+x+d}Ba_{2-x-d}Cu_3O_y + fBaCuO_2 \qquad (3)$$

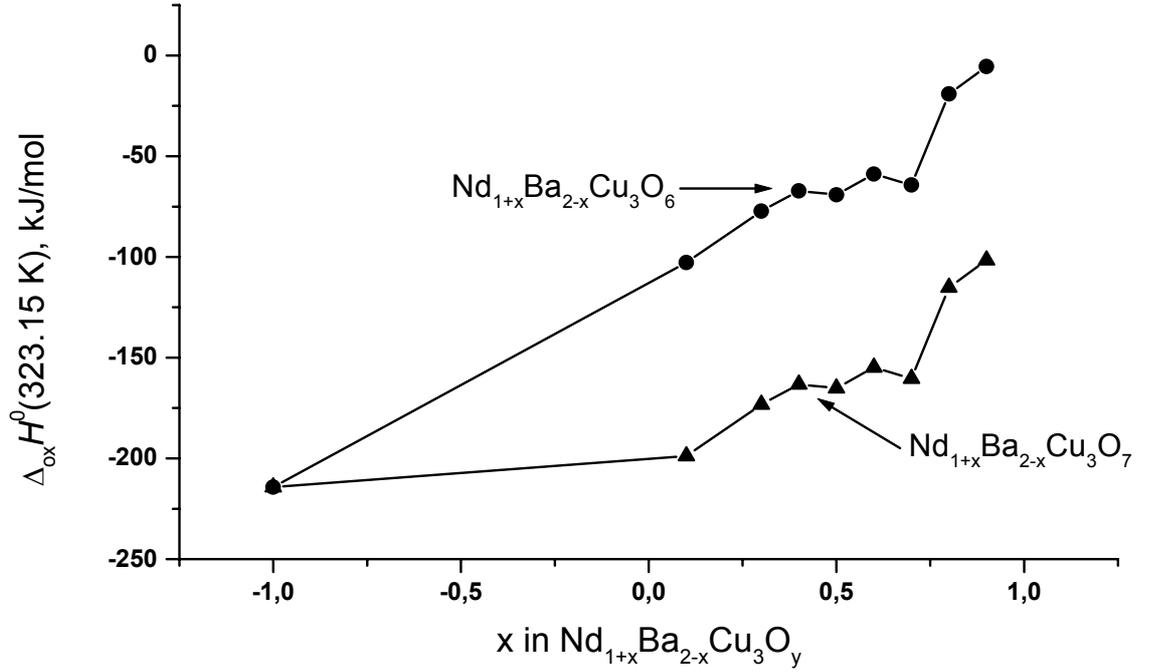

Fig. 1. Dependences of formation enthalpies of $Nd_{1+x}Ba_{2-x}Cu_3O_y$ solid solution from Nd content

In oxygen atmosphere the additional reactions can take place:

$$jNd_{1+x}Ba_{2-x}Cu_3O_y + kO_2 = lNd_{1+x+d}Ba_{2-x-d}Cu_3O_{y+\delta} + mNd_{1+x-d}Ba_{2-x+d}Cu_3O_{y+\delta} \quad (4)$$

$$pNd_{1+x}Ba_{2-x}Cu_3O_y + qO_2 = sNd_{1+x+d}Ba_{2-x-d}Cu_3O_{y+\delta} + tBaCuO_2 \quad (5)$$

Thus, the decomposition both in inert and in oxygen atmosphere may occur in the Nd-Ba-Cu-O system.

Analysis of obtained thermochemical data allowed us to assume the following. First of all, Nd123ss with increasing and decreasing content of Nd can be formed at treatment of solid solutions at high temperature. Then it is possible to form $BaCuO_2$. The existence of decomposition of solid solutions is checked in papers [12, 14, 15]. The formation of different solid solutions should lead to decreasing $T_c$ but to increasing the critical current density because the decomposition leads to appearance of defects. It results in increasing pinning centers and increasing critical current. This assumption can be checked experimentally.

We will analyze data on transport characteristics of single crystals. The highest critical current density ($10^5$ A/sm$^2$) was shown for samples, which were processed as following. It has been grown as describe above. They were oxidized in three steps: 1) at 550° C for 91 h; 2) at 600° C for 23 h; 3) at 550-350° C for 560 h in 1 bar $O_2$. Then the samples were subjected to high-pressure oxidation in three steps: 1) at 500° C for 2 h in 140 bar; 2) at 450° C for 91 h in 140 bar; 3) at 400° C for 96 h in 140 bar $O_2$. But these samples showed a low $T_c$ value about 90.5 K. We explain this low $T_c$ value and high current density by a weak spinodal decomposition which has taken place either during cooling down the grown crystals under an oxygen pressure which has not been low enough or during the 1 bar oxidation in the temperature range 550-600° C.

So, obtained thermochemical data are in an agreement with results of measurements of transport characteristics.

**Acknowledgments**

This work is supported by Russian Fund of Fundamental Investigation (Project No 02-03-32514) and Grant of President of Russian Federation for Supporting of Leading Scientific Scholl (SC-1042.2003.3).

**Conclusion**

1. Thermochemical characteristics of $Nd_{1+x}Ba_{2-x}Cu_3O_y$ solid solutions were measured using solution calorimetry. Dependences of formation enthalpies from Nd content were constructed.

2. It was established that decomposition of solid solution on the basis of Nd123 phase took place both in inert and in oxygen atmosphere. As a result of decomposition different mixtures such as solid solutions with increasing and decreasing Nd content and $BaCuO_2$ were formed.

3. Obtained thermochemical data were compared with results of measurements of transport characteristics. It was established that treatment of samples at 550-600° C led to increasing critical current and decreasing $T_c$. This result was in an agreement with data on decomposition of solid solutions obtained by calorimetry.

**References**

1. Th. Wolf, A.-C. Bornarel, H. Kupfer, R. Meier-Hirmer, B. Obst, Phys. Rev. B 56 (1997) 6308.

2. M. Murakami, N. Sakai, T. Higuchi, S.I. Yoo, Supercond. Sci. Technol. 9 (1996) 1015.

3. M. Nakamura, Y. Yamada, T. Hirayama, Y. Ikuhara, Y. Shiohara, S. Tanaka, Physica C 259 (1996) 295.

4. K. Osamura, W. Zhang, Z. Metallkd. 84 (1993) 522.

5. D. V. Peryshkov, E.A. Goodilin, M.V. Makarova, E.A. Trofimenko, S.N. Mydretsova, A.F. Maiorova, Yu.D. Tretyakov, Doklady of Akademii Nauk. 383 (2002) 651.

6. N.I. Matskevich, E.Yu. Prohorova, O.V. Prokuda, G. Krabbes, Th. Hopfinger, Th. Wolf, Electronic J. "Investigated in Russia" 173 (2002) 1919.

7. M. Murakami, S.I. Yoo, T. Higuchi, N. Sakai, M. Watahiki, N. Koshizuka, S. Tanaka, Physica C 225 (1994) 2781.

8. L.F. Cohen, A.D. Caplin, A.A. Zhukov, G.K. Perkins, J.T. Totty, J.V. Thomas, V.I. Voronkova, D.E. Lacey, Supercond. Science and Technology 12 (1999) 135.

9. G. Krabbes, Th. Hopfinger, C. Wende, P. Diko, G. Fuchs, Supercond. Sci. Technol. 15 (2002) 665.

10. Th. Wolf, H. Kupfer, and P. Schweiss, Physica C 341-348 (2000) 1347.

11. N.I. Matskevich, E.A. Trofimenko, Yu.D. Tretyakov, Chemistry for substantial development 10 (2002) 765.

12. V.V. Petrikin, E.A. Goodilin, J. Hester, E.A. Trofimenko, M. Kakihana, N.N. Oleinikov, Yu.D. Tretyakov, Physica C 340 (2000) 16.

13. N.I. Matskevich, F.A. Kuznetsov, D. Feil, K.-J. Range, Thermochim. Acta 319 (1998) 1.

14. M. Nakamura, Y. Yamada, T. Hirayama, Y. Ikuhara, Y. Shiohara, S. Tanaka, Physica C 295 (1996) 259.

15. G. Krabbes, W. Bierger, P. Schatzle, G. Fuchs, J. Thomas, Adv. Solid State Phys. 383 (1999) 39.